\documentclass[10pt,letterpaper]{article}
\usepackage[top=0.85in,left=2.75in,footskip=0.75in]{geometry}

\usepackage{changepage}

\usepackage[utf8]{inputenc}

\usepackage{textcomp,marvosym}

\usepackage{graphicx}

\usepackage{fixltx2e}

\usepackage{amsmath,amssymb}
\usepackage{gensymb}

\usepackage{cite}

\usepackage{nameref,hyperref}

\usepackage{microtype}
\DisableLigatures[f]{encoding = *, family = * }
\raggedright
\usepackage[aboveskip=1pt,labelfont=bf,labelsep=period,justification=raggedright,singlelinecheck=off]{caption}

\makeatletter
\renewcommand{\@biblabel}[1]{\quad#1.}
\makeatother

\date{}

\usepackage{lastpage,fancyhdr,graphicx}
\usepackage{epstopdf}
\fancyheadoffset[L]{2.25in}
\fancyfootoffset[L]{2.25in}

\begin{document}
\vspace*{0.35in}

\begin{flushleft}
{\huge
\textbf \newline{Spatial dynamics of synthetic microbial 
\\  hypercycles and their parasites}
}
\newline
\\
Daniel R. Amor\textsuperscript{1,2,3},
Ra\'ul Monta\~{n}ez\textsuperscript{2,3,4},
Salva Duran-Nebreda\textsuperscript{2,3},
Ricard Sol\'e\textsuperscript{2,3,5*}
\\
{\small
\bigskip
{1} Physics of Living Systems, Department of Physics, Massachusetts Institute of Technology, 02139 Cambridge, Massachusetts, United States of America.
\\
{2} ICREA-Complex Systems Lab, Department of Experimental and Health Sciences, Universitat Pompeu Fabra, 08003 Barcelona, Spain.
\\
{3} Institute of Evolutionary Biology (CSIC-Universitat Pompeu Fabra), Passeig Mar\'itim de la Barceloneta 37, 08003 Barcelona, Spain.
\\
{4} Centre for Biomedical Network Research on Rare Diseases (ISCIII). U741, 29071 M\'alaga, Spain
\\
{5} Santa Fe Institute, 1399 Hyde Park Road, Santa Fe NM 87501, USA}
\\
\bigskip

* ricard.sole@upf.edu

\end{flushleft}
\section*{Abstract}
Mutualisms are pivotal interactions that shaped some of the major key evolutionary innovations, such as the emergence of eukaryotic cells. Early theoretical work revealed that the simplest class of autocatalytic cycles, known as hypercycles, provide an elegant framework for understanding the evolution of mutualism. Furthermore, hypercycles are highly susceptible to parasites, spatial structure constituting a key protection against them. However, there is an insufficient experimental validation of these theoretical predictions, in addition to little knowledge on how environmental conditions could shape the spatial dynamics of hypercycles. Here, we constructed spatially extended hypercycles by using synthetic biology as a way to design mutualistic and parasitic {\em E. coli} strains. A mathematical model of the hypercycle front expansion is developed, providing analytic estimates of front speed propagation. Moreover, we explore how the environment affects the mutualistic consortium during range expansions. Interestingly, moderate improvements in environmental conditions (namely, increasing the availability of growth-limiting amino acids) can lead to a slowing-down of the front speed. Our agent-based simulations suggest that opportunistic depletion of environmental amino acids can lead to subsequent high fractions of stagnant cells at the front, and thus to the slow-down of the front speed. Moreover, environmental deterioration can also shape the interaction of the parasitic strain towards the hypercycle. On the one hand, the parasite is excluded from the population during range expansions in which the two species mutualism can thrive (in agreement with a classical theoretical prediction). On the other hand, environmental deterioration (e.g., associated with toxic chemicals) can lead to the survival of the parasitic strain, while reshaping the interactions within the three-species. The evolutionary and ecological implications for the design of synthetic consortia are outlined.

\section*{Author Summary}

In order to achieve greater levels of complexity, complex systems often display cooperative interactions that enable the formation and stabilisation of mutualisms. Theoretical models have shown that hypercycles, i. e. positive feedback circuits where different species (or molecules) are helped and help other species by forming closed loops, might be a key mechanism. However, parasites can easily destroy the cooperative loop, unless the system is embedded in a spatial context where interactions are limited to nearest neighbours. Here we explore this problem by engineering synthetic cooperative strains of microbes that grow and interact in a cell culture under the absence and presence of a synthetic parasitic strains. By analysing the impact of cooperation under different conditions, we find that hypercyclic replication is successful and overcomes competitive interactions in nutrient-poor environments. However, the hypercycle fails to establish in nutrient-rich media. Moreover, parasitic entities that jeopardise cooperation under well-mixed conditions can be overcome by hypercycles when growing in a two-dimensional space.


\section*{Introduction}
The evolution of complexity is largely grounded in the emergence of new forms of cooperation capable of holding together higher-order entities from simpler ones. Cooperative interactions have played a great role in the so-called {\em major transitions in evolution} \cite{MSmith1995}. Cooperation pervades the rise of molecular systems capable of overcoming mutation thresholds, multicellular assemblies incorporating division of labour or the appearance of insect societies. Each of these structures incorporates new properties that cannot be observed at the level of its component parts. Despite the burden involved in sustaining the new, larger entity, the advantage of staying together can overcome, under some circumstances, the cost of the association. 

Cooperation can be achieved in particular by means of closed catalytic loops. Mutualistic interactions pervade ecological communities at many different scales, from bacterial communities to microbiomes and large-scale ecosystems \cite{Bronstein2015}. The presence of these reciprocal relations was already outlined by Charles Darwin in one of his memorable studies on the ecology of earthworms \cite{Darwin1892,Wilkinson2006} and summarised by the diagram of fig 1a. Here earthworms improve soil porosity and organic content that helps plants to grow, which results in more organic matter and mechanisms of soil preservation (which favours the earthworm population). This is a simple, two-component ($n=2$) diagram, but ecosystems are characterised by the presence of multiple feedback loops and thus interactions might be more complex, like the three-member ($n=3$) loop shown in (fig 1b). Here vegetation is grazed by animals, whose activity enhances the survival of invertebrates, which in turn improve soil quality thus favouring plant growth. Because of their ecological and evolutionary relevance, cooperative interactions have also been a major topic in synthetic biology \cite{Momeni2011, Hosoda2011,MomeniDemixing,MomeniCheater,Celiker2013,Shou2007}. The possibility of engineering {\em de novo} cooperative interactions is of relevance for several reasons. On one hand, engineered mutualisms could be used to build desirable (even optimal) functionalities that require the presence of a tight metabolic dependence \cite{Wintermute2010, Agapakis2011}. 
Moreover, the possibility of designing mutualistic interactions and even symbiotic pairs \cite{Kiers2011,Guan2013,Hom2014,Agapakis2011,Alvarez2015}
provides a unique opportunity for exploring the emergence of cooperation in evolution under a `synthetic" perspective \cite{Sole2016}.

Mutualistic interactions are also required to sustain stable communities, particularly when harsh conditions are present. An example (fig 1c) is provided by drylands \cite{Weber2016} and in particular the interactions between the so-called biological soil crust (BSC) and vascular plants \cite{Maestre2016}. The BSC defines in itself a complex ecosystem enclosed within a few centimetres of the topsoil, largely controlling the energy and matter flow through the soil surface, helping vegetation thrive under semiarid conditions. The soil microbiome plays a major role in sustaining plant diversity and its dynamics, with the latter often completely dependent on their microbial symbionts \cite{Heijden2008}. Since these ecosystems might experience sudden declines due to climate change \cite{Folke2004,Sole2007} understanding their dynamics is crucial to predicting their future. In this context, it has been suggested that engineering new synthetic mutualistic loops in endangered ecosystems could help prevent catastrophic shifts \cite{SoleTERRA2015,SoleetalTERRA2015}.

\begin{figure}[t]
        \begin{center}
	\includegraphics[width=0.99 \textwidth]{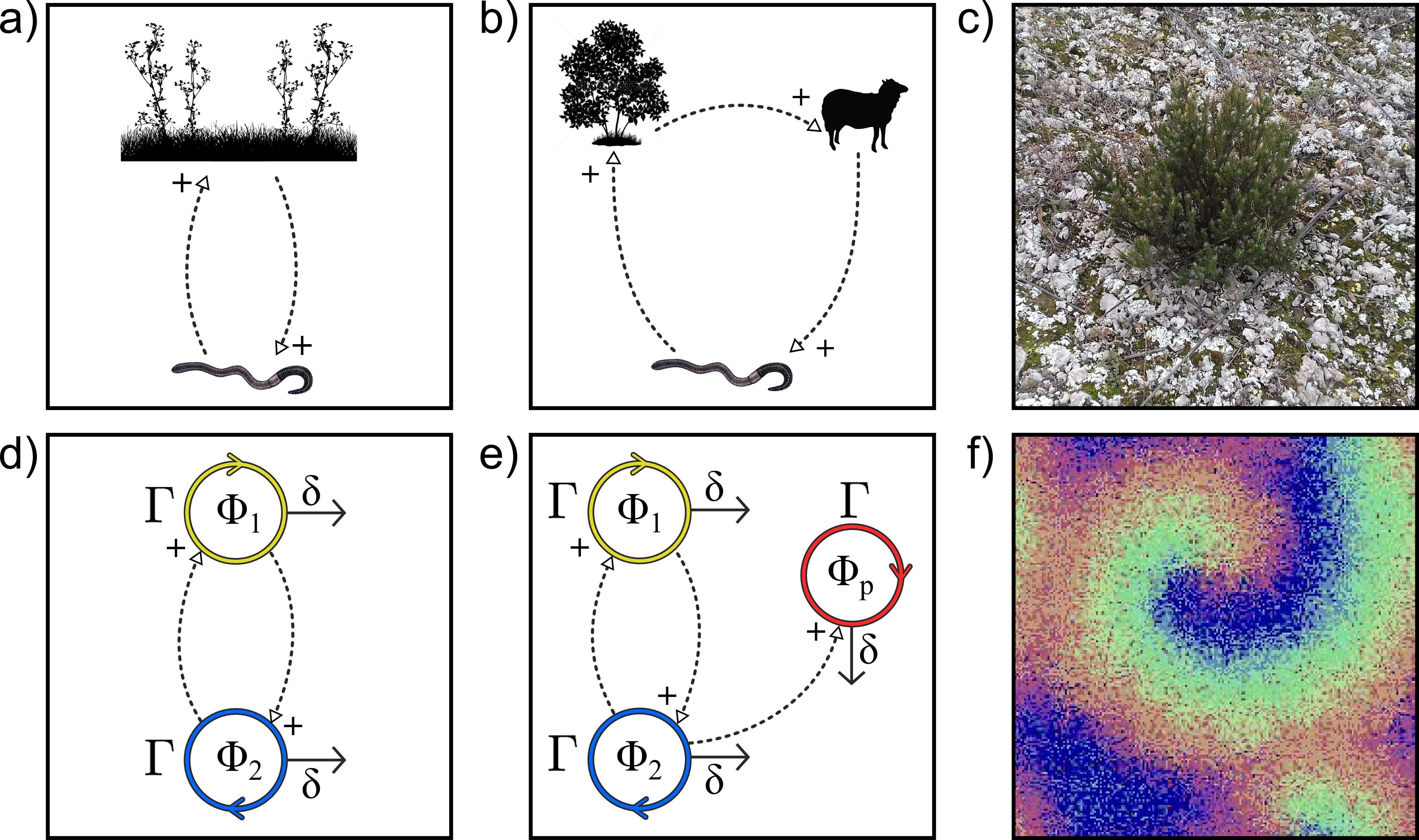}
	\vspace{0.3 cm}
        \caption{\textbf{Spatial structure helps hypercycles to thrive against parasites}. Hypercycles are widespread in ecological systems, and three examples are shown in (a-c). Here we indicate in (a) the mutual support between vegetation (grasses) and earthworms and in (b) a more complex hypercycle composed by vegetation, cattle and earth worms (and other invertebrates). In (c) the image shows a small area within a semiarid ecosystem including a plant surrounded by biological soil crust. Formal models of these types of interactions are described by hypercycles. In (d) we display the basic logical scheme of interactions for a two-member hypercycle.  In (e) we show an extended model where a parasitic species (colour circle) takes advantage of one of the species but gives no mutual feedback.  
        These parasites can easily destroy the hypercycle, but this effect is reduced or suppressed under the presence of oscillations and spatial diffusion when spiral waves get formed (f). Here different colours indicate different molecular species in a $n=8$ member hypercycle (adapted from Attolini and Stadler 2011).}
        \end{center}
\end{figure}

Understanding cooperation, its rise and fall and how can it overcome competitive interactions is an important problem. A great insight has been obtained from both field and theoretical studies \cite{Bronstein2015}. An elegant description of this class of cooperative loops is the {\em hypercycle}, first suggested within the context of prebiotic evolution \cite{Eigen1978,Kauffman,EorsReplicators2006,Higgs2015,Schuster2016}. Here a simple catalytic system is defined (as in figs 1a-b) forming a closed graph where the replication of each component is catalysed by a previous one in the loop, while it also catalyses the replication of the next. The simplest case is the one shown in fig 1d for a two-member syste \cite{Eigen1978,SardanyesSole2006}.  If we indicate by $\Phi_1$ and $\Phi_2$ their population sizes, a pair of coupled equations allows us to represent the hypercycle model as follows: 

\begin{equation}
 \label{Hc1}
  \begin{aligned}
   {d\Phi_1 \over dt} = \alpha_{12}\Phi_1 \Phi_2 \left (1 - \frac{\Phi_1 +\Phi_2}{K} \right ) - \delta_1 \Phi_1\\
   {d\Phi_2 \over dt} = \alpha_{21}\Phi_1 \Phi_2 \left  (1 - \frac{\Phi_1 +\Phi_2}{K} \right ) - \delta_2 \Phi_2 		
 \end{aligned}
\end{equation}

\noindent
where $K$ stands for the carrying capacity of the system, $\delta$ is the degradation/death rate of both species and the replication rates of the cross-catalytic loop are indicated by $\alpha_{ij}$.  As defined, we can see that no proliferation of any of the two partners will occur in the absence of the other, as a consequence of the second-order kinetics that requires the product of the two concentrations.  

The hypercycle can outcompete other non-cooperative species \cite{Eigen1978,EorsReplicators2006} but a major drawback is that it can also be easily threatened by a parasite (figure 1e) capable of destabilising the whole system \cite{MSmith1979}. Interestingly, mathematical and computer models indicate that this problem can be limited by the presence of diffusion in a spatial domain \cite{Boerjlist1991,Boerjlist1994,Attolini2006,JosepRic2007}. 
Hypercycles displaying spatial structures (fig. 1f) are obtained from $n>4$ loops capable of exhibiting oscillations. In a nutshell, the spatial structure imposes a limitation to the spread of the parasite, and it can even go extinct if the inaccessibility of its target species, combined with its death rate, makes it non-viable. This suggestion has received considerable attention \cite{May1991}. Although these models are reasonable for molecular systems, real populations of microbial mutualists spread in space as expanding fronts that impose particular constraints which can be analysed using front propagation theory \cite{Hengeveld1989,Shigesada1997}. It is in this context that previous theories on hypercycle propagation and parasitic interactions can be tested using synthetic ecologies.  

In this paper, we address this problem by using an experimental design where populations of synthetic microbes are forced to cooperate as they expand on a two-dimensional substrate. In this context, we take into account the population spreading that leads to complex spatial structures, partially due to the cooperative loop but also to the physical impact of cellular shapes. Both populations had some potential for (Malthusian) growth in the absence of the mutualistic partner, provided that the necessary metabolites are present. Such response allows testing the conditions under which the hypercycle overcomes the effects of competition derived from Malthusian populations exploiting similar resources. As will be shown below, the contact surface between the engineered strains increases with the strength of mutualistic interactions. When the mutualistic interactions are neutralised, the synthetic strains display the same segregative dynamics described by competing invaders \cite{Hallatscheck2007,Hallatscheck2009}. Moreover, a synthetic parasite was also designed to test the capacity of the spatial synthetic hypercycle to prevent it from spreading. Additionally, our parasitic strain has been engineered to degrade ampicillin from the medium, in order to explore the boundaries between parasitic and cooperative interactions,

\section*{Results}

\begin{figure}[tp]
\begin{center}
\includegraphics[width=0.98 \textwidth]{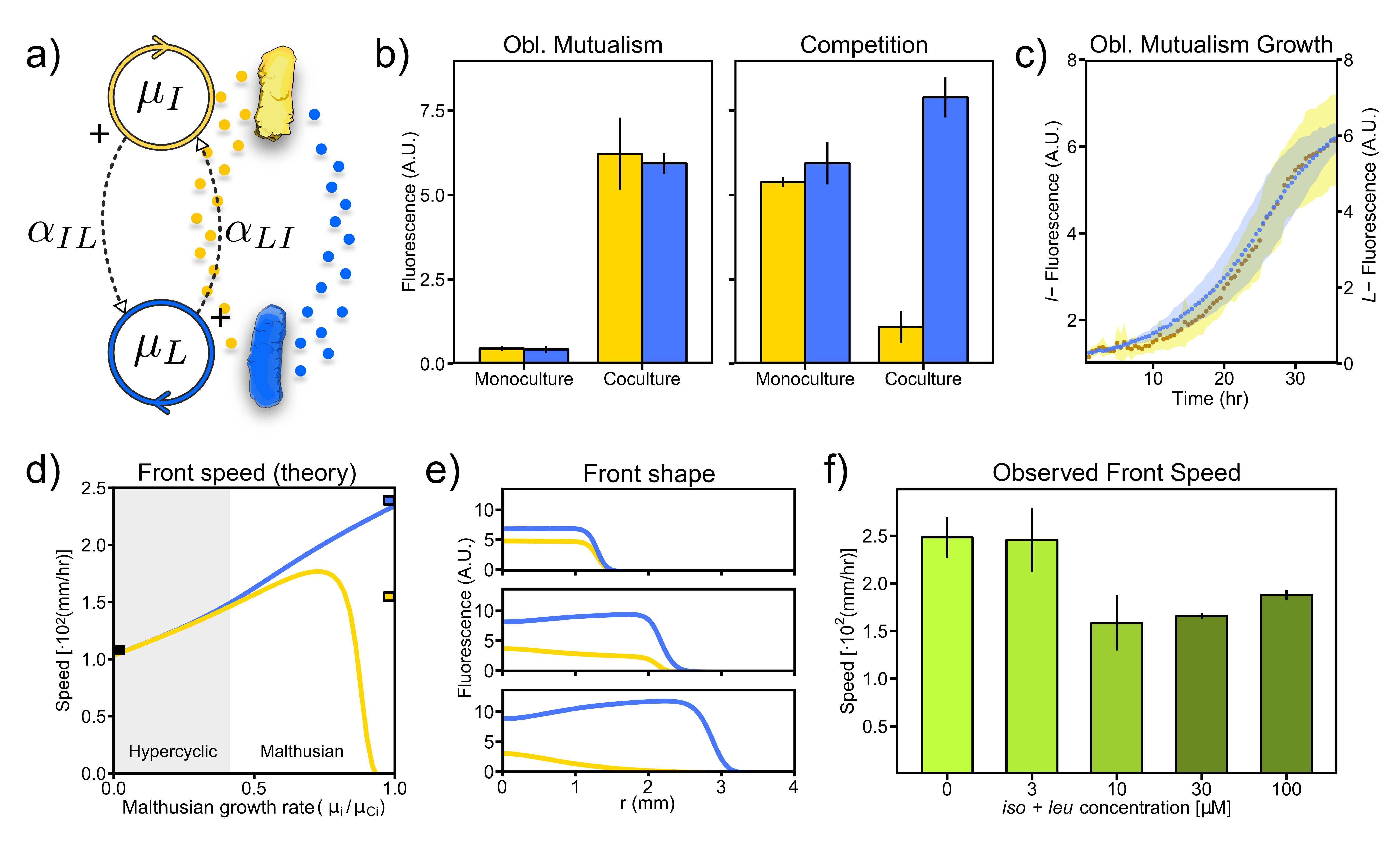}
\caption{{\bf Malthusian versus hypercyclic dynamics in synthetic consortia} \textit{a)} We use a pair of engineered bacterial strains (yellow depicts \textit{I\textsuperscript{ -}} cells and blue stands for  \textit{L\textsuperscript{-}}) that engage in mutualistic interactions by cross-feeding amino acids. \textit{b)} Both strains are able to grow cooperatively in liquid cultures lacking both amino acids, but monocultures exhibit no growth in this conditions (Obl. Mutualism). When amino acids are supplemented at $10^{-4}$M (Competition), monocultures grow to comparable levels while the \textit{L\textsuperscript{-}} strain overcomes its partner in cocultures. \textit{c)} Time series (dots stand for average, shades for standard deviation) showing the coupled growth of the (obligate) hypercycle in liquid medium lacking both \textit{iso} and \textit{leu}. 
 \textit{d)}  Invasion speed of the mutualistic strains according to a minimal reaction-diffusion model. The gray area indicates the domain where the cooperative interaction favours hypercyclic growth over 
 Malthusian competition (parameter values extracted from observed growth for \textit{I\textsuperscript{ -}} and \textit{L\textsuperscript{-}} in liquid cultures, see Supp. Info.). \textit{e)} Front shape for the two strains for different self-reproduction rates (which models the effect of supplemented amino acids in the medium). The top panel shows the obligate ($\mu_{i}=0$) hypercycle case: the coupled populations propagate as two travelling waves that approximately share the location of their fronts' edges. In the medium panel ($\mu_{i}=\mu_{Ci}/2$), the two species display interactions at the critical intersection that separate mutualism from competition: both strains travel at similar speeds, but the front edge of \textit{I\textsuperscript{-}} remains slightly behind one of \textit{L\textsuperscript{-}} due to its smaller growth rate in the presence of amino acids. In the lower panel ($\mu_{i}=\mu_{Ci}$), the faster replicator \textit{L\textsuperscript{-}} wins the competition by conquering the available space long before \textit{I\textsuperscript{-}}, which is progressively let behind until it is excluded from the population range expansion process. \textit{f)} Observed front speed for cocultures spreading on agar surfaces, exhibiting a slowing-down in facultative mutualism scenarios that are not captured by the RD model.}
\label{RD}
\end{center}
\end{figure}

Our model system to study mutualistic interactions is composed of the pair of bacterial engineered strains shown in Fig. \ref{RD}a. The \textit{I\textsuperscript{ -}} strain (depicted in yellow) cannot produce the isoleucine (\textit{iso}) amino acid but overproduces and leaks leucine (\textit{leu}), while \textit{L\textsuperscript{-}} (in blue) cannot produce \textit{leu} but overproduces and leaks \textit{iso}\cite{Hosoda2011}. Therefore, the strains are able to engage in a cross-feeding mutualism that permits growth in coculture, in a minimal medium lacking both amino acids where neither \textit{I\textsuperscript{ -}} nor \textit{L\textsuperscript{-}} can grow in monoculture (obligate mutualism scenario in Fig. \ref{RD}b). However, both \textit{I\textsuperscript{ -}} and \textit{L\textsuperscript{-}} are able to grow in monoculture when this same medium is supplemented with $10^{-4}$M of both \textit{iso} and \textit{leu} (competition scenario in Fig. \ref{RD}b, see also \nameref{S1_Fig} and \nameref{S2_Fig}).

\subsection*{Malthusian growth can break-down the hypercycle during range expansions}

How the synthetic hypercycle would spread in different environments is a focal question we address here. We used different concentrations of supplemented amino acids in order to modify the interactions displayed by our synthetic pair \textit{I\textsuperscript{ -}} and \textit{L\textsuperscript{ -}} (as done in Refs. \cite{Muller2014, Hoek2016} for mutualistic yeast strains). Regarding the hypercycle minimal model, let us consider amino acid supplementation as a way to introduce Malthusian growth rates for the species in the system (see the observed Malthusian growth rates in \nameref{S1_Fig} for the competition scenario). Furthermore, to be able to explore growth on solid culture conditions, the simple model [Eq. set (\ref{Hc1})], must be redefined to incorporate space. To this aim, let us consider the following Reaction-Diffusion (RD) approach \cite{Murray2004, Korolev} as a minimal model describing the spatiotemporal dynamics of the synthetic mutualistic replicators:

\begin{equation}
 \label{RDEq}
  \begin{aligned}
   \frac{\partial I}{\partial t}=D\frac{\partial ^{2}I}{\partial r^{\ 2}}+(\mu_{I}I+\alpha_{IL}IL) \left (1-\frac{I+L}{k} \right),\\
   \frac{\partial L}{\partial t}=D\frac{\partial ^{2}L}{\partial r^{\ 2}}+(\mu_{L}L+\alpha_{LI}IL) \left (1-\frac{I+L}{k} \right)
  \end{aligned}
\end{equation}

\noindent
where $I$ and $L$ stand for the population density of the \textit{I\textsuperscript{ -}} and \textit{L\textsuperscript{-}} strain respectively, $t$ and $r$ are the time and spatial coordinates (see Methods), $D$ is the diffusion coefficient, $\mu_i$ is the Malthusian growth rate of species $i$, $\alpha_{ij}$ ($\geq 0$) is the growth rate of species $i$ assisted by its mutualistic partner $j$, and $k$ is the carrying capacity of the system. The above set of equations generalised the two-member hypercycle model by including, on the one hand, the spatial context (through the diffusion terms $D\partial^2 /\partial r^2$) and, on the other, by considering both Malthusian ($\mu_k\geq 0$) and mutualistic ($\alpha_{ij} \geq 0$) growth terms.

By considering the absence of either species in the set (\ref{RDEq}), we recover the one-species Fisher RD model \cite{Fisher1937, Skellman} that leads to the well-known expression for the invasion speed:


\begin{equation}
 \begin{aligned}
  c_{I_{F}}=2\sqrt{\mu_{I}D}	\qquad\text{for}\qquad L=0, \\
  c_{L_{F}}=2\sqrt{\mu_{L}D}	\qquad\text{for}\qquad I=0
  \end{aligned}
 \label{cF}
\end{equation}
 
Moreover, the Fisher speed establishes the asymptotic invasion speed for our two-species system in(\ref{RDEq}) as $\mu_i >> \alpha_{ij}$ (for $i=I,L$ and $i\neq j=I,L$). In the case of two purely competing species ($\mu_{i}>0$, and $\alpha_{ij}=0$) we should expect the front to propagate at the speed of the faster competitor because this species will be more efficient at conquering the available space at the edge of the population front. In contrast, for the case of two purely mutualistic species (i.e., a pure hypercycle with $\mu_{i}=0$, and $\alpha_{ij}>0$), we derived the analytical solution for the invasion speed (see Methods): 

\begin{equation}
c=\sqrt{\frac{Dk\alpha_{IL}\alpha_{LI}}{2(\alpha_{IL}+\alpha_{LI})}}
\label{Hc}
\end{equation}

Our minimal model (\ref{RDEq}) thereby predicts two different invasion modes for our pair of mutualistic strains \textit{I\textsuperscript{ -}} and \textit{L\textsuperscript{-}}. Indeed, in the competition scenario, the invasion speed (\ref{cF}) is governed by the growth rate at low population densities(which gives rise to a pulled front [REF]). In contrast, the carrying capacity $k$ appearing in Eq. (\ref{Hc}) is a hallmark of an invasion front governed by the growth dynamics at high population densities. This gives rise to a pushed front \cite{Gandhi} : individuals at the edge of the front are pushed from the inside bulk where individuals reproduce at higher rates. Moreover, note that the invasion speed (\ref{Hc}) is the same for the two mutualists \textit{I\textsuperscript{ -}} and \textit{L\textsuperscript{-}}, consistent with their need for a mutualistic partner in order to grow and spread.

Figure \ref{RD}d shows how the transition between the two invasion modes takes place, according to the RD model. In the absence of Malthusian replication ($\mu_i=0$), both strains spread at the same speed. As both $\mu_I$ and $\mu_L$ are increased towards their observed value (see \nameref{S1_Fig}) in the competition scenario, the front speed increases due to the corresponding enhancement in growth rates. However, once $\mu_i$ induces stronger effects on the front than $\alpha_{ij}$, competition becomes important and the coupled advance of the two strains is replaced by two differentiated front speeds. At this point, further increasing the Malthusian growth rates $\mu_i$ benefits the faster species (in this case, the \textit{L\textsuperscript{-}} strain), while the second one is slowed down in a relatively abrupt way. This eventually leads the \textit{I\textsuperscript{ -}} strain to be excluded from the front (which propagates at the Fisher's speed $c_{L_{F}}$ as Malthusian growth rates approach the observed values in competition). The population density profiles in Fig \ref{RD}e illustrate the change in the invasion front shape as the scenario transits from obligate mutualism to competition.

Self-reproduction rates of the auxotrophic strains can be experimentally tuned by supplementing the medium with different doses of the two amino acids (an analogous method was used in Ref \cite{Muller2014} for yeast mutualists).Fig. \ref{RD}f shows that the observed speeds for cocultures spreading on agar (See Methods) are not consistent with those observed in \ref{RD}d, within the considered range of amino acid concentrations. The particularly low values of the front speed observed in the experimental transition, from the obligate mutualism to the competition scenario, revealed one of the limitations of the minimal RD model. According to the RD model, the minimum front speed predicted for the synthetic hypercycle should be the invasion speed of the obligate mutualists. In other words, even if one of the strains is slowed down because of competition, the edge of the front will keep travelling at the speed of the fastest strain (which should exceed the speed of the obligate hypercycle in order to overcome its partner species at the edge of the front). Thus, the decrease of the observed front speed as supplemented amino acids are increased indicates that other, more complex phenomena are driving the dynamics of the synthetic hypercycle. In particular, the physical embodiment of bacterial cells (not taken into account by the RD model) may affect their access to the extracellular amino acids, thus influencing the invasion speed.
 

\subsection*{Slowdown of hypercycle front speed under local resource depletion in moderately rich environments}

\begin{figure}[t]
 \includegraphics[width=0.9 \textwidth]{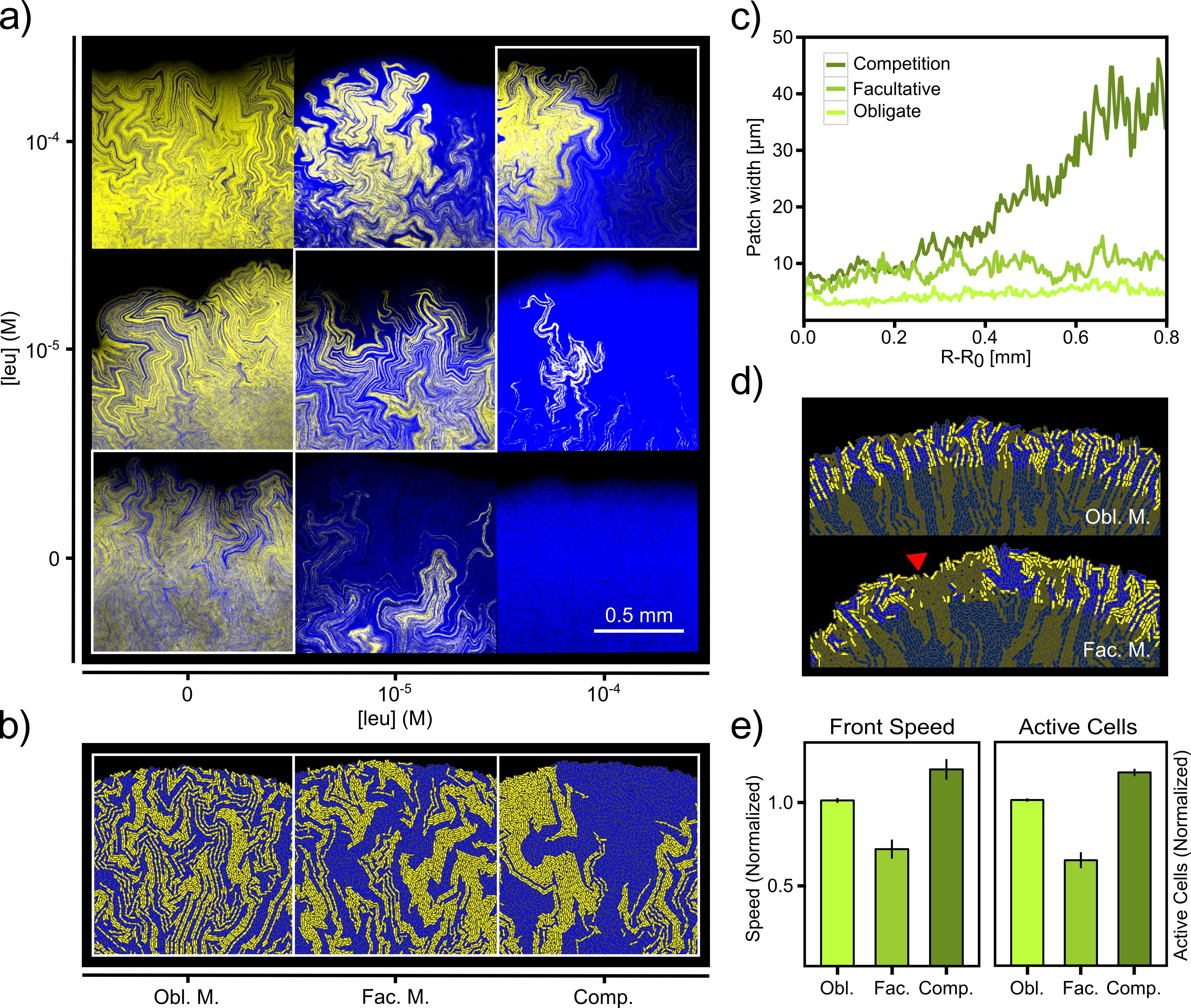}
  \caption{{\bf Facultative mutualism can slow down the front of synthetic hypercycles} \textit{a)} Spatial structure close to the edge of the population front after four days of incubation. Different concentrations of supplemented \textit{iso} and \textit{leu} lead to different spatial dynamics at the edge of the front (e.g., $[iso]=0$ and $leu=10^{-4}$M leads the \textit{L\textsuperscript{-}} strain to govern the front). White rectangles indicate the obligate mutualism, facultative mutualism and competition scenarios. \textit{b)} Snapshots of agent-based simulations reproducing the three main scenarios. \textit{c)} Average width of single-strain patches from experimental data. \textit{d)} Snapshots of simulated fronts (darker colours depict stagnant cells). The red arrow indicates a patch of \textit{I\textsuperscript{ -}} cells formed by local consumption of environmental amino acids. Once amino acids are locally depleted, a high number of cells in the patch become stagnant. \textit{e)} Normalised speed and normalised active cells in three simulated scenarios.}
 \label{Pattern}
\end{figure}

In order to further study the role of spatial structure in range expansions from synthetic hypercycles and how it is influenced by the strength of mutualistic interactions, we seed the cross-feeding system on M63 plates with 1.2$\%$ agar and different concentrations of auxotrophic amino acids distributed as is depicted in figure \ref{Pattern}a. Each point in the experimental setting represents a different value of $\alpha_{ij}$ \cite{Muller2014,Hoek2016}. When no amino acids are supplemented into the medium, cells are only able to growth if mutualistic partners remain close enough, the population engages in an obligate mutualism, which leads to a self-organized distribution with a characteristic high intermixing of the two strains. The opposite scenario (\textit{iso} and \textit{leu} at $10^{-4}$M) in Fig. \ref{Pattern}a reveals a remarkably different spatial structure. When the driving interaction is competition for space and resources, the invasion dynamics is governed by genetic drift \cite{Hallatscheck2007}, which leads to demixing of the population into wide (single-strain) patches. This experimental condition correlates with $\mu_{Mi}$ in the RD minimal model (\ref{Hc}), the competition scenario. In these conditions, we would expect the slower replicator (lower $\mu_i$) to go extinct at the edge of the advancing front (see the competition case in Fig \ref{RD}e). However, in our experimental scenario (Fig. \ref{Pattern}a, top-right panel), both strains are present at the edge of the front despite exhibiting significantly different growth rates $\mu_I < \mu_L$ (see Supp Info). This result is consistent with the expected effects of genetic drift in population range expansions \cite{Hallatscheck2009}.

In between of the above two modes of invasion, we found the environmental conditions that allow a facultative mutualistic behaviour. Single-strain patches are wider than those observed in the absence of supplemented \textit{iso} and \textit{leu}, although genetic diversity is still preserved (the characteristic width of patches is preserved) as the front propagates, Fig. \ref{Pattern}c. In other words, in the facultative scenario, the concentration of amino acids added to the media permit the strains to grow into wider patches (compared to those of obligate mutualists), but both strains still benefit from the cross-feeding. However, in the competition scenario, the high concentration of amino acids permits the two strains to spread at comparable speeds regardless of the presence of the mutualistic partner (see Supp Info). Once mutualism is suppressed, genetic drift becomes the governing mechanism at the edge of the front, leading to progressively wider (\ref{Pattern}c) single-strain patches as the front advances.

The scenarios in Fig \ref{Pattern}a reveal a qualitatively identical interplay between mutualism and genetic drift in range expansions of yeast populations Ref. \cite{Muller2014}, which suggests that such feature is universal and independent of the biological organisms exhibiting the mutualism. However, boundary domains between the yeast strains in Ref. \cite{Muller2014} are significantly different than the ones in Fig \ref{Pattern}a. The differences on the boundary domains could be explained by the different cell shape \cite{Rudge2013}. In order to develop an agent-based model for range expansions of the bacterial hypercycle that consider cell shape, we used the {\em GRO} package \cite{Klavins2012}. Our agent-based model takes into account the cross-feeding interactions between the two strains and captures the experimentally observed scenarios at the edge of the front as a function of the supplemented amino acid concentrations (see Fig. \ref{Pattern}b). Moreover, the model also captures different boundary domains associated to cell shape (as we show in \nameref{S3_Fig}). 

GRO simulations allow us to study whether local depletion of supplemented amino acids could effectively slow-down the invasion process similarly as observed in experimental conditions. Nutrients and amino acids are mainly consumed by cells at the edge of the front \ref{Pattern}d, their depletion leaves a population of stagnant cells that effectively constitutes a fossil record of the invasion process \cite{Hallatscheck2007}. In the obligate mutualism case, single-strain patches keep a characteristic width determined by the distance at which cells can sustain the cross-feeding mutualism (cells near the front can temporarily become stagnant when their location prevents an effective cross-feeding). This process shapes the spatial distribution of the population, leading to a relatively high fraction of active cells at the edge of the front (\ref{Pattern}d and e). However, in the case of facultative mutualism, the dynamics can be marked by episodes of opportunistic growth that exploits the available amino acids in the environment. During these periods, the dynamics are locally governed by genetic drift (single-strain sectors become wider). However, once the supplemented amino acids are locally depleted, a significant number of cells (remote to the boundary domains where cross-feeding is still effective) can become stagnant (arrow in Fig \ref{Pattern}d). Figure \ref{Pattern}e shows how the ratio of active cells is correlated with the invasion speed, suggesting that the dynamics in facultative mutualism scenarios can slow-down the invasion speed of the synthetic hypercycle.

\begin{figure}[tp]
 \includegraphics[width=0.9 \textwidth]{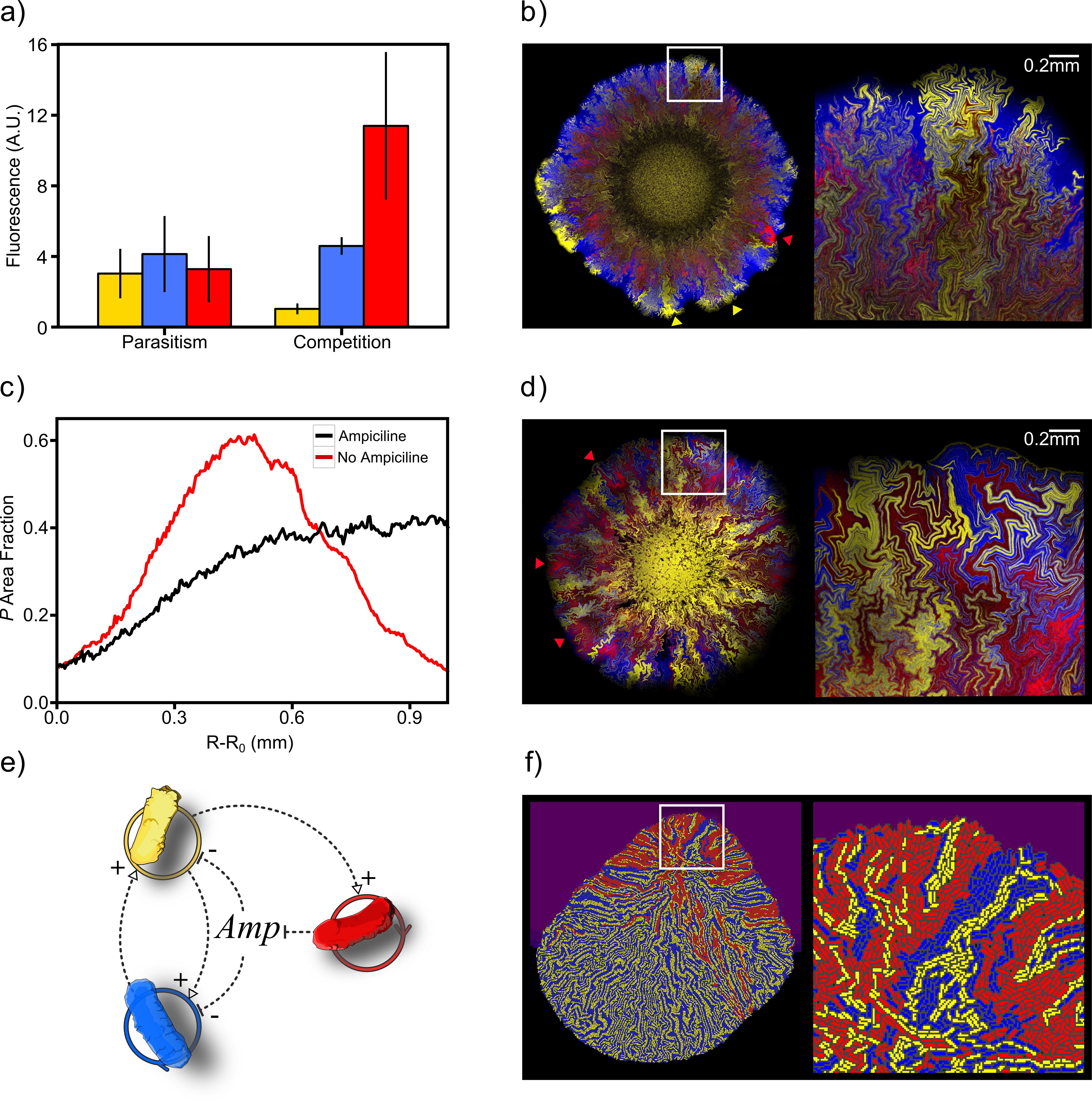}
  \caption{{\bf Environmental conditions determine the fate of parasites during range expansions.} \textit{a)} Obligate mutualism scenario (absence of supplemented amino acid)  leads the strain \textit{P} to act as a parasite in well-mixed conditions, while competition is observed at $10^{-4}$M supplementation of \textit{iso} and \textit{leu}. \textit{b)} Spatial structure leads the mutualists to conquer the edge of the population front, defeating the parasite \textit{P}. Yellow arrows indicate regions where the parasite has been excluded from the population front (red arrow indicates one of the few regions in which the parasite still surf at the population wave). Note that the front curvature is enhanced at regions governed by the mutualists, a hallmark of an enhancement of the front speed at these regions. The grey rectangle indicates the magnified area on the right. \textit{c)} Frequency of the \textit{P} strain at the edge of the front for two different scenarios ($0$ and $100\mu$M extracellular ampicillin). \textit{d)} The \textit{P} strain offers cross-protection to the mutualists when threatened by antibiotics, leading to the survival of the \textit{P} strain at the edge of the front. \textit{e)} Scheme of the complex mutualistic interaction (which involves cross-feeding and cross-protection) between the three species in the presence of antibiotics. Each species lacks a different ability needed to survive in the system, but the ensemble may be able to survive if able to develop the corresponding division of labour. \textit{f)} Three-species spatial structure in a simulated heterogeneous environment with non-isotropic antibiotic concentration at $t=0$. While the \textit{P} strain is conserved in the areas where cross-protection is essential for the mutualistic ensemble,  \textit{P} cells are excluded from the front in areas where the antibiotic concentration does not reach the growth inhibition threshold.}
\label{Cheater}
\end{figure}

\subsection*{Environmental deterioration can determine the survival of parasites during range expansions}

Several processes (such as mutations or the arrival of foreign, invader species) may give rise to new organisms exploiting hypercycle feedbacks in a given ecosystem. The introduction of a new replicator organism that makes use of the limited resources in the medium will restrict the growth of the hypercycle, even more, if this new organism is a parasite (hereafter \textit{P} cells), that takes advantage of the cross-feeding (Fig. \ref{Cheater}e).

In order to experimentally study the ecological implications of such parasites, we used the synthetic parasitic strain \textit{P} (see Methods) that exploits one of the cross-feeding amino acids (namely, \textit{iso}).
The coculture of those three organisms in well-mixed conditions, for both the obligate mutualism and the competition scenarios, give as result a restricted growth of \textit{I\textsuperscript{ -}} or \textit{L\textsuperscript{-}} strains, Fig. \ref{Cheater}a shows (compare to Fig \ref{RD}b). Moreover, for the competition scenario in Fig \ref{Cheater}a, the \textit{P} strain exhibits a relatively high Malthusian growth rate (see Supp Info) that leads it to overcome the growth of the hypercycle pair. 

To test whether spatial structure can limit the parasitic exploitation of hypercycles, we coculture combinations of the three strains (\textit{I\textsuperscript{ -}}, \textit{L\textsuperscript{-}} and \textit{P}) on M63-agar plates. In the absence of supplemented amino acids, when \textit{I\textsuperscript{ -}} or \textit{L\textsuperscript{-}} cells are lacking, no growth was observed. This means that \textit{P} cells can be considered a hypercycle parasite, because they are unable to close an effective cross-feeding loop (see Fig. \ref{RD}a) with either \textit{I\textsuperscript{ -}} or \textit{L\textsuperscript{-}} cells. When the three strains are present (Fig. \ref{Cheater}b), despite an initial success of the parasite at colonising available space (see fig. \ref{Cheater}c, red line), the parasitic strain is progressively left behind as the range expansion takes place. This is because, in the spatial scenario, cell location determines a preferential access to the cross-feeding metabolites\cite{MomeniCheater,Pande2016}. Thereby, the presence of a \textit{P} patch increases the distance between \textit{I\textsuperscript{ -}} and \textit{L\textsuperscript{-}} and leads to restricted growth. This gives a significant advantage to mutualistic \textit{I\textsuperscript{ -}} and \textit{L\textsuperscript{-}} neighbouring patches that engage in an efficient cross-feeding. Hence, spatial structure benefits the hypercycle species, eventually leading the hypercycle ensemble to overcome the parasite at the edge of the front (Fig \ref{Cheater}b).

The ecological role of a species in a given community can be strongly dependent on its environment and transitions can occur between mutualism and parasitism as external conditions change \cite{Bronstein2015,Bronstein1994,Herre1999,Hernandez1998,Neuhauser2004}. In our three-member microbial consortium, composed by \textit{I\textsuperscript{ -}} \textit{L\textsuperscript{-}} and \textit{P}, we studied whether environmental deterioration can make this community to develop a more complex mutualistic network. In order to do this,  the three-member microbial consortium was seeded on m63-agar plates containing a lethal concentration of ampicillin, for which \textit{P} cells are resistant. The \textit{P} cells are able to degrade extracellular Ampicillin (by secreting beta-lactamase). Now, two different mutualistic motives are present in this scheme (Fig \ref{Cheater}e): (amino acids) cross-feeding and (antibiotic) cross-protection. Remarkably, the hypercycle trio was able to solve the complex environmental problem and develop the range expansion process on the corresponding agar layers. Figure \ref{Cheater}d shows the observed spatial structure displayed by this new mutualistic ensemble in order to collectively invade the available space. In contrast to the previous parasitic case, the fraction of the \textit{P} strain is approximately constant as the population front advances (see Fig \ref{Cheater}c).

The definition of the three-member consortium as an agent-based model allows us to make some predictions on how the system would spread within heterogeneous environments and captures the main spatial dynamics features of the system (see Supp Info). Simulation in a heterogeneous environment, that presents an asymmetric spatial antibiotic distribution, allows us to see how the \textit{P} strain remains present at the edge of the front in the top region of the colony, which is precisely where the population is exposed to higher doses of antibiotic. In contrast, in the lower region where the antibiotic dose is much lower, the \textit{P} strain is excluded from the edge of the front (consistently with our previous results), (Fig. \ref{Cheater}f).

This is an interesting result particularly within the context of bioengineering soils \cite{SoleTERRA2015,SoleetalTERRA2015} by the rewiring of the ecological interactions within the biological soil crust (BSC). Here the vertical structure defines a heterogeneous set of conditions where different species and physicochemical spatial gradients are present. Both in the BSC and around the plant root system a complex microbiome exists. Soil engineering under a systems perspective is a promising domain to harness and restore different functionalities \cite{Dejong2010}. This approach could be complemented by designed microbiomes exploiting mutualistic ties following some of the basic findings reported here. Since different soil conditions might sustain different qualitative functional traits, the previous synthetic three-species ecosystem can inspire novel forms of improving soil communities and plant efficiency.

\section*{Discussion}

Most experimental and theoretical studies concerning the dynamics of microbial populations are grounded in competition. However, cooperation is a crucial component of ecological dynamics on all scales, and much needed to truly understand the behaviour of a wide range of systems, from populations growing on biofilms to the gut microbiome. Moreover, it has been suggested that synthetic cooperation can help to design ecological circuits capable of preventing endangered ecosystems from collapsing \cite{SoleTERRA2015,SoleetalTERRA2015}. 

Previous studies have analysed a family of models involving closed cooperative loops of cooperators. These systems are known as hypercycles, and because of their second-order kinetics, they are capable of hyperbolic growth. This kind of dynamics allows the hypercycle to overcome the simple Malthusian replicators. Theoretical work shows that hypercycles can prevent their decay due to the presence of parasites by exploiting the constraints imposed by a spatially extended system. However, these models require some special properties concerning the nonlinear dynamics of hypercyclic sets, which are not feasible in realistic conditions.
Instead, we have analysed the problem of hypercycle persistence and response to parasites by means of experimental setups where populations of engineered cooperators spread on a two-dimensional medium. 

Our study reveals that, as predicted by theoretical models involving both linear (Malthusian) growth and hypercyclic cooperation, spatial dynamics makes a big difference when space is introduced as propagating fronts. This is favoured by both the microscopic impact of bacterial shapes (leading to characteristic fractal structures) and by the local correlations required to sustain cooperation, which favour a maximisation of contact domains between the two cell populations. Hypercyclic growth has been characterised using diverse sets of measures and the front speed mathematically derived from a diffusion model. 

The experiments and models confirm the picture of spatial hypercycles as dynamical systems where the mutualistic tie forces the formations of complex structures that guarantee the propagation of the cooperative consortium. We have also studied the tradeoffs associated with Malthusian growth and the conditions pervading the breakdown of hypercyclic cooperation thus showing the presence of two phases: one associated with competitive interactions and a second phase associated with scarce resources promoting the cooperative feedback. 

The second set of experiments and models are related to the impact of parasitic strains on the stability of the hypercycle. Parasites are easy to evolve under standard conditions of growth in cell cultures: when synthetic constructs have been added to microbial strains, the loss of one construct has an immediate impact on the metabolic burden, thus allowing the cell to replicate faster. However, the presence of a given dependency can create nontrivial interactions. We designed synthetic parasitic strains capable of exploiting a given amino acid while not completing the cooperation cycle. Such parasite (which has a small component of Malthusian growth) has been shown to overcome and kill the hypercycle under liquid conditions but becomes a much less harmful component under spatial constraints. It was recently shown that resource availability can modulate the interactions between microbial cross-feeding mutualists\cite{Hoek2016}. Our work is, as far as we know, the first experimental design of a synthetic ecological network showing how different contexts allow cooperation, competition or parasitism to succeed or even transition from one to the other in a spatially extended context. Further work should explore how these results translate into more realistic contexts, from the gut microbiome to soil ecosystems.

\section*{Materials and Methods}

\subsection{Theoretical invasion speed of a 2-species hypercycle}

Our theoretical RD model for the two-species hypercycle considers that the dynamics of the system is governed by diffusion and population growth as:

\begin{equation}
 \label{RMD}
  \begin{aligned}
   \frac{\partial I}{\partial t}=D\frac{\partial ^{2}I}{\partial r^{\ 2}}+(\mu_{I}I+\alpha_{IL}IL) \left (1-\frac{I+L}{k} \right),\\
   \frac{\partial L}{\partial t}=D\frac{\partial ^{2}L}{\partial r^{\ 2}}+(\mu_{L}L+\alpha_{LI}IL) \left (1-\frac{I+L}{k} \right)
  \end{aligned}
\end{equation}
For convenience, we rewrite this set of equations in terms of dimensionless variables $I^*=I/k$, $L^*=L/k$, $t^*=\alpha_{IL}kt$ and $r^*=(\alpha_{IL}k/D)^{1/2}r$, and dimensionless parameters $\alpha^{*}=\alpha_{LI}k/\alpha_{IL}$. Then, the new set reads:

\begin{eqnarray}
\frac{dI^*}{dt^*}=\frac{\partial ^{2}I^*}{\partial r^{*2}}+I^*L^* (1-I^*-L^*) \\
\frac{dL^*}{dt^*}=\frac{\partial ^{2}I^*}{\partial r^{*2}}+\alpha^*I^*L^* (1-I^*-L*), 
\label{MRDim}
\end{eqnarray}

Let us assume that there exist travelling wave-shaped solutions of the previous equations of the form:
\begin{eqnarray}
I^*(r^*,t^*)=U_I(z)=\xi_I \frac{1}{(1+ae^{bz})^s}\\
L^*(r^*,t^*)=U_L(z)=\xi_L \frac{1}{(1+ae^{bz})^s},
\label{sol1}
\end{eqnarray}
with $s>0$, $b>0$, $a>0$, and $z=r-ct$ (where $c$ is the speed of the travelling wave, i.e. the front speed of the hypercyclic population). Using
$$\frac{dU_i}{dx}=\frac{dU_i}{dz}=U_i'$$
$$\frac{dU_i}{dt}=-c\frac{dU_i}{dz}=cU_i'$$ 
with $i=I,L$, the set (\ref{MRDim}) can be rewritten as:

\begin{equation}
U_I''+cU_I'+U_IU_L(1-U_I-U_L)=0
\label{MRDI}
\end{equation}

\begin{equation}
U_L''+cU_L'+\alpha^*U_IU_L(1-U_I-U_L)=0,
\label{MRDL}
\end{equation}

Developing the derivatives $U_I''$ and $U_I'$, Eq. (\ref{MRDI}) reads:

\begin{equation}
\begin{split}
\varepsilon_I  [s(s+1)\eta^{-s-2}a^2b^2e^{2bz}-s\eta^{-s-1}ab^2e^{bz}\\
-sc\eta^{-s-1}abe^{bz}\\
+\varepsilon_L\eta^{-2s}-\varepsilon_I\varepsilon_L\eta^{-3s}-\varepsilon_{L}^{2}\eta^{-3s}]=0, 
\end{split}
\label{D1}
\end{equation}

where $\eta=(1+ae^{bz})$. Neglecting the trivial solution ($\varepsilon_I=0$) for Eq. (\ref{D1}), and reorganising terms according to powers of $e^{bz}$, we obtain the characteristic equation for the front speed $c$:

\begin{equation}
\begin{split}
e^{2bz}[s(s+1)a^2b^2]\\
+e^{bz}[-sa\eta(b^2+bc)]\\
+\varepsilon_L\eta^{-s+2}+\varepsilon_I\varepsilon_L\eta^{-2s+2}+\varepsilon_L^2\eta^{-2s+2}=0
\end{split}
\label{D2}
\end{equation}

Solutions for the travelling wave have to be valid $\forall z$, and thus each line in Eq. (\ref{D2}) gives an independent expression that must necessarily vanish. Analysing the terms in the last line in Eq. (\ref{D2}) leads to the necessary condition $s<2$. This leads to $s=1$ because we only consider solutions with $s>0$. Then, considering $s=1$, we develop the conditions given by the different powers of $e^{bz}$ in Eq. (\ref{D2}), which leads to:

\begin{equation}
\varepsilon_I=1-\varepsilon_L,
\label{D3}
\end{equation}

\begin{equation}
c=\frac{\varepsilon_L-b^2}{b},
\label{D4}
\end{equation}
and 
\begin{equation}
b=c.
\label{D5}
\end{equation}

Combining Eqs. (\ref{D3})-(\ref{D5}) leads to:
\begin{equation}
c=\sqrt{\varepsilon_L/2}=\sqrt{(1-\varepsilon_I)/2}.
\label{D6}
\end{equation}

With an analogous procedure to the one performed above for Eq. (\ref{MRDI}), analysis of Eq. (\ref{MRDL}) leads to:
\begin{equation}
c=\sqrt{\alpha^*\varepsilon_I/2}=\sqrt{\alpha^*(1-\varepsilon_L)/2}.
\label{D7}
\end{equation}

Combining Eqs. (\ref{D6}) and (\ref{D7}) we obtain the expressions for the species abundances in the travelling front:
\begin{equation}
\begin{split}
\varepsilon_I=1/(1+\alpha^*), \\
\varepsilon_L=\alpha^*/(1+\alpha^*)
\end{split}
\label{D8}
\end{equation}

Replacing terms from Eq. (\ref{D8}) into Eq. (\ref{D7}), we obtain the analytical solution for the front speed in dimensionless variables:
\begin{equation}
c=\sqrt{\frac{\alpha^*}{2(1+\alpha^*)}}.
\label{D9}
\end{equation}

Finally, recovering dimension variables, the speed of the front reads:

\begin{equation}
v=c\sqrt{Dk\alpha_{IL}}=\sqrt{\frac{Dk\alpha_{IL}\alpha_{LI}}{2(\alpha_{IL}+\alpha_{LI})}} 
\label{D10}
\end{equation}

\subsection*{The agent based model}

Our approach to the study of hypercycles reveals the importance of considering cells as embodied 
entities, both as interacting elements on a microscopic scale and as spatially extended populations. 
Moreover, cells need to incorporate the molecular circuits associated to the specific regulatory 
mechanisms along with chemical reactions, spatial diffusion and molecular signalling. 
To this goal, we used the specification language {\em gro} \cite{Klavins2012}
as the platform for individual-based simulation of growing populations. 

Our model integrates the main physical features of bacterial shape and growth \cite{Klavins2012}, as well as the cross-feeding and cross-protection interaction between \textit{I\textsuperscript{ -}} \textit{L\textsuperscript{-}} and \textit{P} strains. We used a very simple approach that considers a few step (Heavyside) functions to emulate cell behaviour. A list of the considered cell behaviour features follows:

\begin{enumerate}
\item Sensing: at each time step, each cell senses the extracellular concentration of three kinds of molecules: amino acids (\textit{I\textsuperscript{ -}} cells sense \textit{iso}, while \textit{L\textsuperscript{-}} and \textit{P} cells sense \textit{leu}), food (this category embraces any other nutrients that cells may need to grow),  and antibiotic (i.e., ampicillin).

\item Growth: cells grow (increase their cell volume) and divide at the realistic speed proposed in Ref. \cite{Klavins2012}, provided that:
\begin{enumerate}
\item food concentration exceeds a given threshold value $g_f$.
\item the corresponding amino acid (according to cell strain) exceeds a given threshold value $g_{am}$.
\item antibiotic concentration is below a given inhibitory threshold $g_{at}$.
\end{enumerate}
Accordingly, cell growth is arrested whenever any of the above conditions are violated.

\item Cells absorb extracellular food and release amino acid (or $\beta$-lactamase) at constant rates, provided that extracellular food exceeds $g_f$. Specifically, \textit{I\textsuperscript{ -}} cells release \textit{leu}, \textit{L\textsuperscript{-}} cells release \textit{iso}, and \textit{P} cells release the betalactamase enzime (that degrades the antibiotic) to the extracellular medium. Provided that growth conditions are satisfied, cells will also absorb the amino acid they need. 

\end{enumerate}

\begin{figure}[tp]
	\begin{center}
                \includegraphics[width=0.7 \textwidth]{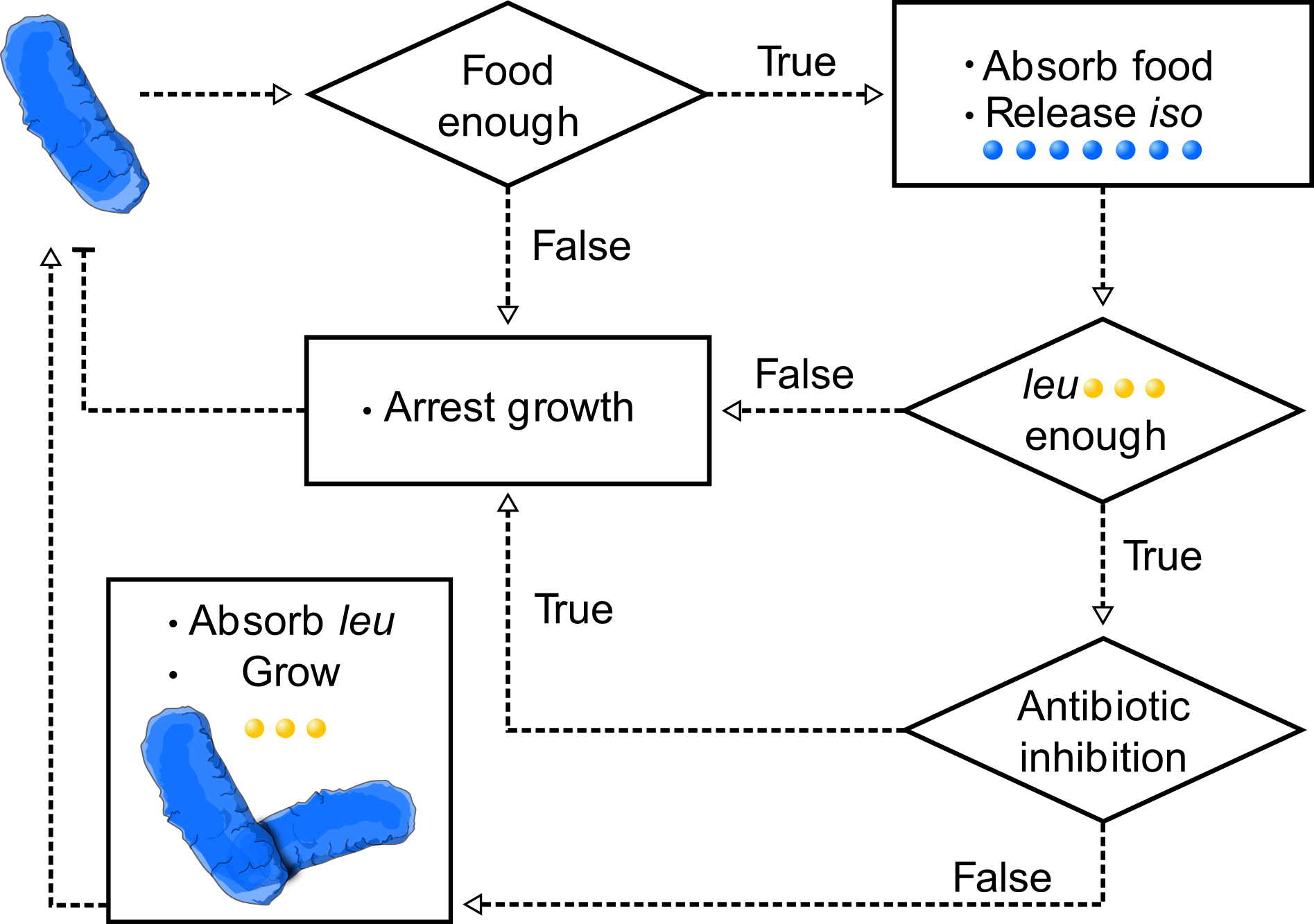}
\caption{{\bf Cell dynamics in the agent-based model is governed by binary decisions (Heaviside behaviour) that depend on extracellular concentration thresholds of nutrients, amino acids and antibiotics.}}
\label{Logical}
	\end{center}
\end{figure}

The corresponding logical loop experienced by a given \textit{L\textsuperscript{-}} cell at each time step is illustrated in Fig \ref{Logical}. \textit{I\textsuperscript{ -}} and \textit{P} cell dynamics follow analogous logical schemes.

Furthermore, in order to consider a fitter parasitic strain that evades the cost of the mutualism in antibiotic-free scenarios, we consider the growth rate of \textit{P} cells to be higher (by a $10\%$ difference) than that of \textit{I\textsuperscript{ -}} and \textit{L\textsuperscript{-}} cells. As shown above, the hypercycle was able to escape the parasite despite such faster growth rate.

Admittedly, actual cell dynamics is far more complex than this Heavyside representation. However, our goal for the agent-based model was to use a minimal set of assumptions, in order to provide an easy understanding of the key features governing the system dynamics. Remarkably, the Heavyside-based cell behaviour is enough to capture the essential dynamics, as discussed in the Results section. The source code and additional details on specific values for metabolic rates and concentration threshold values can be found in the Supp. Info.

\subsection*{Bacterial strains}

Both the \textit{I\textsuperscript{ -}} and the \textit{L\textsuperscript{-}} strains are from E. coli strain DH1 (National BioResource Project, National Institute of Genetics, Shizuoka, Japan) and were genetically modified to cross-feed as described in \cite{Hosoda2011}. The \textit{I\textsuperscript{ -}} (\textit{L\textsuperscript{-}}) strain carries the \textit{dsred.T3 (gfpuv5)} gene that provides the corresponding fluorescence labelling.

 Cloning for the \textit{P} strain was carried out using the Biobrick assembly method and the parts: B0014, J23100, B0032 and E0020, from the Spring 2010 iGEM distribution assembled into a low copy number plasmid pSB4A5. A complete description of the construction protocols can be found at \cite{Carbonell-Ballesteros_2014, Carbonell-Ballesteros_2016}.
  
\subsection*{Culture conditions}

All regular cultures and amplifications were done at 37 $\celsius$ in well-mixed media Lysogeny Broth (LB). Bacterial strains were cryopreserved in LB-glycerol 20$\%$ (v/v) at -80 $\celsius$. Along experiments, cells were grown at 37 $\celsius$ in well-mixed minimal media modified with M63 (pH 7.0, 62 mM $K_{2}HPO_{4}$, 39 mM $KH_{2}PO_{4}$, 15 mM ammonium sulfate, 1.8 $\mu$M $FeSO_{4}-7H_{2}O$, 15 $\mu$M thiamine hydrochloride, 0.2 mM $MgSO_{4}-7H_{2}O$ and 22 mM glucose; mM63 \cite{Kashiwagi_2009}).

For individual cloning selection, Leucine\textsuperscript{ -} and Isoleucine\textsuperscript{ -} cells taken from glycerinated were grown overnight 24h in LB at 37 \celsius, diluted and plated on Petri dishes with M63 agar (1.2$\%$ agar), $10^{-4}$ M of the auxotrophic amino acid and selected with the appropriate antibiotics (chloramphenicol 30 $\mu$g/ml; kanamycin 20 $\mu$g/ml); The parasitic cells were grown overnight (O/N) in (LB) at 37 \celsius, diluted and plated on Petri dishes with LB agar and kanamycin (25 $\mu$g/ml).

12h before each experiment, 5 colonies of each type of cell were selected and grown separately in LB plus $10^-4$ M of auxotrophic amino acid at 37 \celsius. After 12h, a 100-fold dilution with fresh amino acid supplemented LB plus antibiotics (for auxotrophic cells) or 500-fold (for the parasite), with fresh LB plus cabenicillin (100 $\mu$g/ml), were done and grown until OD660$\sim$0.4.

\subsection*{Fluorescence assays}
Fresh cultures, with a OD660$\sim$0.4, were washed twice by centrifugation and resuspension with M63 medium. A final density of cells of $OD660 nm = 0.3$ for monoculture and $OD660 nm = 0.15 + 0.15$ for cocultures was seeded in a flat bottom 96-well microplate (Sarstedt AG $\&$ Co. Germany), with M63 antibiotics. Growth was monitored over time, by quantification of the fluorescence that identifies each kind of cell (mRFP for L\textsuperscript{ -}, GFP for I\textsuperscript{ -} and CFP for the parasite). M63 without cells was included in the incubation as a background control for both fluorescence and absorbance. Incubation and measures of bacterial cultures during characterization were performed on a Synergy MX-microplate reader (BioTek Instruments, USA) every 30 min for 62 h. Fluorescence measures for RFP (ex: 560$\pm$9 nm, em: 588$\pm$9 nm), GFP (ex: 478$\pm$9 nm, em: 517$\pm$9 nm) and CFP (ex: 450$\pm$9 nm, em: 476$\pm$9 nm) with gain 90 were carried out, as well as optical density (OD at 660 nm) measures. Incubation was done at 37 $\celsius$ with continuous orbital shaking (medium intensity). Leucine and Isoleucine concentration conditions were prepared from an initial stock at 0.5 M and serial dilutions in M63, ranging from 10\textsuperscript{-4} to 10\textsuperscript{-10} M were prepared the day of the mutuality to competition transition experiments.

\subsection*{Range expansions on agar surfaces}
Fresh cultures, with a OD660$\sim$0.4, were washed twice by centrifugation and resuspension with M63 medium. The optical density of each culture was settled to OD660 nm $=$ 0.15. Equitable volumes were mixed to generate cocultures with 2 or 3 different cells. Finally, 0.4 $\mu$L of the mono or coculture volume was seeded on the center of a m63 (1.2$\%$ agar). Cells were grown 4 days at 37 $\celsius$ and humidity 90$\%$. colonies were observed using a Leica TCS SP5 AOBS (inverted) confocal microscope.

\section*{Supporting Information}
%

%

\subsection*{S1 Fig}
\label{S1_Fig}
{\bf Malthusian and hypercycle growth rates for the synthetic strains} Time series for the fluorescence of the \textit{I\textsuperscript{ -}} strain, when cultured in M63 medium supplemented with $100$ $\mu$M of both \textit{iso} and \textit{leu}. Coloured dots stand for the average values across 9 replicates (three technical replicates from each of three biological replicates), shaded area indicates standard deviation. The Malthusian growth rate $\mu_{I}$ was obtained by linear regression (black solid line) to the data during the exponential growth regime (region delimited by the vertical dashed lines), as described in \nameref{S1_Text}. \textit{b)} Malthusian growth rate for the \textit{L\textsuperscript{-}} strain (growth conditions as in a)). \textit{c)} Malthusian growth rate for the \textit{P} strain (growth conditions as in a)). Hyperbolic growth rates $\alpha_{IL}$ and $\alpha_{LI}$ were obtained from the observed growth at low population densities (region between dashed lines), as described in \nameref{S1_Text}. The time series correspond to the growth of both  \textit{I\textsuperscript{ -}} and  \textit{L\textsuperscript{-}} strains in coculture, in M63 medium with no supplemented amino acids.

\subsection*{S1 Text}
\label{S1_Text}
{\bf Growth rates in well-mixed conditions}

In order to measure the Malthusian growth rates associated with competition scenarios, we cultured each of the three strains in M63 medium supplemented with $100$ $\mu$M of both \textit{iso} and \textit{leu} amino acids. Fluorescence measures showed consistency with the expected exponential growth regime, that precedes growth saturation when the population reaches its carrying capacity, as shown in \nameref{S1_Fig}. Malthusian growth rates for the three species were obtained through linear regression of the observed growth data, according to the Malthusian growth model: 

\begin{equation}
log(F)=  \mu_{j}t+\beta ,
\label{S1}
\end{equation} 
 
 where $F$ stands for the fluorescence value, $\mu_{j}$ is the Malthusian growth rate of species $j$, $t$ stands for the time and $\beta$ is a constant value for $t=0$. Thus we obtained the values $\mu_{I}=(9.1 \pm 0.1) \times 10^{-2}$ hr$^{-1}$, $\mu_{L}=(2.18 \pm 0.02) \times 10^{-1}$ hr$^{-1}$ and $\mu_{P}=(3.75 \pm 0.02)\times 10^{-1}$ hr$^{-1}$. 

 In order to obtain an estimate for the hyperbolic growth rates, we considered the well-mixed version of Eq. (2) in the Main text. This correspond to the following set of equations, which do not account for the diffusion process:
 
\begin{equation}
 \label{WellM}
  \begin{aligned}
   \frac{\partial I}{\partial t}=\alpha_{IL}IL \left (1-\frac{I+L}{k} \right),\\
   \frac{\partial L}{\partial t}=\alpha_{LI}IL \left (1-\frac{I+L}{k} \right),
  \end{aligned}
\end{equation}

where we have also assumed that we deal with the obligate mutualism scenario ($\mu_{I}=\mu_{L}=0$). 

Considering an approach for the growth at low population densities, it is easy to obtain a solution for $I(t)$ and $L(t)$. Thus, let us neglect the carrying capacity effects [last term on the right hand side of the set of Eqs. (\ref{WellM})], which leads us to:
 
\begin{equation}
 \label{WellM2}
  \begin{aligned}
   \frac{\partial I}{\partial t}=\alpha_{IL}IL,\\
   \frac{\partial L}{\partial t}=\alpha_{LI}IL
  \end{aligned}
\end{equation}

 The above set (\ref{WellM2}) permits to write $I(t)$ in terms of $L(t)$ as:

\begin{equation}
 \label{WellM3}
   I=\frac{\alpha_{IL}}{\alpha_{LI}}L+\delta,
\end{equation}

with $\delta=I_{0}-\alpha_{IL}L_{0}/ \alpha_{LI}$ (where $I_{0}$ and $L_{0}$ stand for the population density numbers at the initial instant $t_0$). Replacing Eq. (\ref{WellM3}) into (\ref{WellM2}), and solving the corresponding differential equation we get the following solution for $L(t)$:

\begin{equation}
 \label{WellM4}
   L(t)=\frac{\delta}{exp(-\alpha_{LI}\delta(t+\gamma))-\alpha_{IL}/ \alpha_{LI}},
\end{equation}

where
\begin{equation}
 \label{WellM5}
   \gamma=\frac{a}{\alpha_{LI}}ln(L_0)+\frac{b\alpha_{LI}}{\alpha_{IL}}ln\left ( \frac{\alpha_{IL}L_0}{\alpha_{LI}} +\delta \right) - t_0,
\end{equation}

with $a=1/ \delta$ and $b=-a\alpha_{IL}/ \alpha_{LI}^2$.

The above Eqs. (\ref{WellM3}) and  (\ref{WellM5}) constitute a set that we can adjust to the observed growth in well-mixed conditions, with $\alpha_{IL}$ and $\alpha_{LI}$ as the only adjustable parameters. By applying a least squares algorithm to the the fluorescence time series for the growth of the obligate mutualists, we obtained $\alpha_{IL}= (4.4 \pm 0.1) \times 10^{-2} $hr$^{-1}$ and $\alpha_{IL}= (6.2 \pm 0.1) \times 10^{-2}$ hr$^{-1}$, as \nameref{S1_Fig} shows.

 \subsection*{S2 Fig}
\label{S2_Fig}
{\bf Cell concentration scales linearly to fluorescence for the three species}  \textit{a)} Cell concentration in liquid cultures of the \textit{I\textsuperscript{ -}} strain according to their fluorescence. The value of $a$ indicates the slope (in ml$^{-1}$) obtained by linear regression of the data points. \textit{b)} In agreement with cell concentration, optical density also scales linearly to fluorescence for the\textit{I\textsuperscript{ -}} strain.  \textit{c)} and \textit{d)} show the same analysis as in \textit{a)} and , but for the \textit{L\textsuperscript{-}} (while  \textit{e)} and \textit{f)} correspond to analogous results for the \textit{P} strain).

\subsection*{S1 Table}
\label{S1_Table}
{\bf Relevant parameters in the agent-based model} The table shows the main parameters of the agent-based model, as well as the main processes they affect. Unless stated otherwise in the text, the parameter values used in simulations correspond to those in the source code (\nameref{S2_Text}).

\subsection*{S2 Text}
\label{S2_Text}
{\bf Agent based simulations source code} 
Source code used to run our simulations in the GRO package \cite{Klavins2012}. 


\subsection*{S3 Fig}
\label{S3_Fig}
{\bf Cell shape influences mesoscopic boundary domains} \textit{a)} Fractal dimension for the boundaries between \textit{I\textsuperscript{ -}} and \textit{L\textsuperscript{-}} patches in the obligate mutualism scenario. Bars indicate average values, while vertical lines indicate standard deviation from three different simulations. \textit{b)} A snapshot showing the patches of the \textit{I\textsuperscript{ -}} strain (in white), when de division size parameter is set to $2.0$, for a colony with approximately $1.6 \times 10^4$ individuals. \textit{c)} A snapshot showing the patches of the \textit{I\textsuperscript{ -}} strain (in white), when de division size parameter is set to $3.5$, for a colony with approximately $1.6 \times 10^4$ individuals.

\subsection*{S4 Fig}
\label{S4_Fig}
{\bf Agent-based simulations capture the spatial dynamics of hypercycle range expansions} \textit{a)} Agent-based simulations show analogous scenarios to those observed in Fig 3a. Values on the vertical and horizontal axis indicate the parameter values for the initial extracellular concentration of amino acids ($I0$ and $L0$, respectively, see \nameref{S1_Table}). b) Patch width in simulated range expansions, for a different initial extracellular concentration of amino acids (initial nutrient concentration $F0=90$). c) A biological replicate for each of the cases presented in Fig. 3c in the Main Text.

\subsection*{S5 Fig}
\label{S5_Fig}
{\bf Fraction of \textit{P} strain in range expansions} \textit{a)} In silico, fraction of territory colonized by \textit{P} cells in three-species population range expansions. Three different scenarios are shown: no ampicillin ($Ampi0=0.0 $, see  \nameref{S1_Table}), moderate ampicillin concentration ($Ampi0=2.0$), and high ampicillin concentration ($Ampi0=4.0$). \textit{b)} Biological replicate for the two scenarios in Fig. 4c.

 \section{Front speed for one-species hypercycles}
In the Main Text, we have derived an analytical solution for the front speed of two-species hypercycles. For completion, we here present the theoretical speed that would correspond to a one-species hypercycle. Analogous theoretical front speeds for one species mutualistic populations are also described in Refs. \cite{Murray2004, Korolev}.

Let's consider an expanding one-species hypercyclic population ($u$), modelled under the following reaction-diffusion approach:

\begin{equation}
\frac{du}{dt}=D(\frac{\partial^2 u}{\partial x^2})+ru^2(1-u)
\label{u1}
\end{equation}
   
   We are interested on the invasion speed to the hypercycle. As a first approach, we will consider that the front is planar, and then the one-dimensional speed is a good approximation of the speed of the front in two dimensions. We rescale $t$ and $x$ in order to work with dimensionless variables: $t^{*}=rt$, and $x^{*}=x(\frac{r}{D})^{1/2}$.
 
 Then, Eq. (\ref{u1}) becomes:
 
 \begin{equation}
 \frac{d u}{dt^{*}}=\frac{\partial^2 u}{\partial x^{*2}}+u^2(1-u)
\label{u2}
\end{equation}

We look for front propagation solutions, so let us assume that there exist solutions to Eq. (\ref{u2}) with the propagating wave form:
 \begin{equation}
u(x^{*},t^{*})=U(z)=\frac{1}{(1+ae^{bz})^s},
\label{u3}
\end{equation}
 with $b,s>0$ and $z=x^{*}-ct{*}$. Using, $u_{x}=U_{z}=U'$, and $u_{t}=-cU_{z}=cU'$ (where the subscripts denote the corresponding partial derivatives), we obtain the expressions for the following partial derivatives: 
 
 \begin{equation}
u_{x}=-sabe^{bz}(1+ae^{bz})^{-s-1}
\label{u3}
\end{equation}

\begin{equation}
u_{xx}=s(s+1)a^2b^2e^{2bz}(1+ae^{bz})^{-s-2}-sab^2e^{bz}(1+ae^{bz})^{-s-1}
\label{u4}
\end{equation}

\begin{equation}
u_{t}=csabe^{bz}(1+ae^{bz})^{-s-1}
\label{u5}
\end{equation}

We rewrite Eq. (\ref{u2}) as:
\begin{equation}
U''+cU'+u^2(1-U)=0,\quad \forall z
\label{u6}
\end{equation}
 
 Rewriting Eq. (\ref{u6}), and reorganising terms as powers of $e^{bz}$ we obtain:

\begin{equation}
\begin{split}
 e^{2bz}(s(s+1)a^2b^2-sa^2b^2-csa^2b) \\
+e^{bz}(-sab^2-csab) \\
+(1+ae^{bz})^{-s+2}
- (1+ae^{bz})^{-2s+2} = 0
\end{split}
\label{u7}
\end{equation}
 
 The above Eq. (\ref{u7}) has to be equal to zero $\forall z$. Then ,the coefficients of $e^0$, $e^bz$, and $e^2bz$ must all be identical to zero. Taking into account that we look for travelling waves solution of the form (\ref{u6}) (with $s>0$), it is easy to show that the only value that $s$ can take is $s=1$. Using this into Eq. (\ref{u6}) leads to:
 
 \begin{equation}
\begin{split}
 e^{2bz}(a^2b^2-ca^2b) \\
+e^{bz}(-ab^2-cab+a) = 0,
\end{split}
\label{u8}
\end{equation}
 
 which leads to the dimensionless front speed:
 \begin{equation}
c=b=  \sqrt{1/2} 
\label{u8}
\end{equation} 
 
 Recovering dimension variables from the front speed $c$ we obtain:
  which leads to the dimensionless front speed:
 \begin{equation}
v=  \sqrt{rD/2.} 
\label{u9}
\end{equation}

\section*{Acknowledgments}
We would like to thank the members of the complex systems lab for useful discussions. DA specially thanks Eva Garc\' ia-Ramallo and Carlos Rodr\' iguez-Caso for discussions and advice on experimental protocols. 
Our acknowledgements to the group of Nanako Shigesada, for facilitating the \textit{I\textsuperscript{ -}} and \textit{L\textsuperscript{-}} bacterial strains. This study was supported by an European Research Council Advanced Grant (SYNCOM), a MINECO grant FIS2015-67616-P, by Banco Santander through its Santander Universities Global Division, the Secretaria d'Universitats i Recerca del Departament d'Economia i 
Coneixement de la Generalitat de Catalunya. and by the Santa Fe Institute. 


\end{document}